\documentclass[%
 pra
 aip,
 jmp,%
 amsmath,amssymb,
twocolumn,superscriptaddress]{revtex4-2}

\usepackage{graphicx}
\usepackage{dcolumn}
\usepackage{bm}
\usepackage{hyperref}

\usepackage{booktabs}
\usepackage{physics}
\usepackage{xcolor}
\usepackage{multirow}
\usepackage{float}

\newcommand{\vv}[1]{\boldsymbol{\mathbf{#1}}}

\begin{document}

\preprint{APS/123-QED}

\title{Parity and time-reversal symmetry violation in diatomic molecules:\\
LaO, LaS and LuO}

\author{Yuly Chamorro}
\email{y.a.chamorro.mena@rug.nl}
\affiliation{Van Swinderen Institute for Particle Physics and Gravity, University of Groningen, 9747 AG, Groningen, The Netherlands}
\author{Victor Flambaum}
\affiliation{School of Physics, The University of New South Wales, Sydney, New South Wales 2052, Australia}
\affiliation{Johannes Gutenberg-Universität Mainz, 55099 Mainz, Germany}
\author{Ronald F. Garcia Ruiz}
\affiliation{Massachusetts Institute of Technology, Cambridge, MA 02139, USA}
\author{Anastasia Borschevsky}
\affiliation{Van Swinderen Institute for Particle Physics and Gravity, University of Groningen, 9747 AG, Groningen, The Netherlands}
\author{Luk\'a\v{s} F. Pa\v{s}teka}
\affiliation{Van Swinderen Institute for Particle Physics and Gravity, University of Groningen, 9747 AG, Groningen, The Netherlands}
\affiliation{Department of Physical and Theoretical Chemistry, Faculty of Natural Sciences, Comenius University, Ilkovičova 6, 84215 Bratislava, Slovakia}

\date{\today}

\begin{abstract}

The violation of parity (P) and time-reversal (T) symmetry is enhanced in the LaS, LaO and LuO molecules due to the existence of states of opposite parity with small energy differences and the presence of heavy nuclei. We calculate the molecular enhancement for the P, T-violating electron electric dipole moment ($W_{\mathrm{d}}$), scalar-pseudoscalar nucleon-electron interaction ($W_{\mathrm{s}}$), nuclear magnetic quadrupole moment ($W_{\mathrm{M}}$), and for the nuclear spin-dependent P-violating anapole moment ($W_{\mathrm{A}}$). We use the relativistic 4-components coupled cluster method and perform a systematic study to estimate the associated uncertainties in our approach. We find that the individual contribution of each computational parameter to the total uncertainty in a system is approximately the same for all the calculated enhancement factors, summing up to a total uncertainty of $\sim7$\%. We discuss the energy shifts and matrix elements associated with the calculated molecular enhancement factors and relate them to higher-energy P- and P, T- violating interactions.

\end{abstract}

\maketitle

\section{Introduction}

To explain the origin of the observed asymmetry in the matter-antimatter composition of the Universe is one of the major challenges in modern physics \cite{CanDreSha12}. Sakharov has shown that violation of the CP invariance (where C stands for charge conjugation and P for parity) is one of the necessary conditions for generating this asymmetry \cite{Sak67}. However, the currently observed amount of CP violation is too small to account for the observed matter dominance \cite{GavHerOrl94,aai19}. This discrepancy motivates searches for additional sources of CP violation (or time reversal violation,  which is equivalent assuming CPT conservation). Alongside high-energy experiments, precision measurements on atoms and molecules provide a promising route to search for these effects and for testing the Standard Model of particle physics \cite{SafBudDeM18,Arr24,BoeWil23}. In particular, diatomic and small polyatomic molecules benefit from a high sensitivity to P-, T-violating effects due to various enhancement mechanisms \cite{SusFla78,FlaKhr85,DzuFla12,SafBudDeM18,ChaBorEli22,OleSkrLeo22,DenHaaTimm19,Gar2020}. Many such experiments are focused on the search for the P, T-violating electron electric dipole moment (eEDM), using various paramagnetic molecules, where this effect is enhanced due to their electronic structure \cite{BoeMarMei23,WuHanCho20,AndVitHut18,aggarwal2018measuring,hudson2011improved,loh2013precision,mitra2020direct,nakhate2019pure}. Such experiments are also sensitive to another P, T-violating effect, namely the scalar-pseudoscalar nucleon-electron neutral current interactions (S-PS) \cite{PosRit14}. 

In paramagnetic systems with non-zero nuclear spin $I$, the internal CP-violating nuclear interactions lead to nuclear spin-dependent P, T-violating effects. For $I\geq1/2$ there is a contribution of the nuclear Schiff moment.  However, a significantly bigger P, T-violating effect in paramagnetic systems may be produced by the magnetic interaction of the electrons with the nuclear magnetic quadrupole moment (NMQM) which appears for nuclear spin $I \geq 1$ \cite{SusFlaKhr84,flambaum1994spin}. 

The corresponding P, T-violating effect
can serve as an indicator of CP violation in the hadron sector \cite{SusFlaKhr84,khriplovich1976bound,flambaum1994spin} and can be also used to search for dark matter \cite{stadnik2014axion}.
In heavily deformed nuclei, the NMQM can be significantly enhanced due to collective effects \cite{GinFla04,flambaum1994spin}, and hence paramagnetic molecules containing such nuclei are promising systems for measurements \cite{FlaDeMKoz14}.  

Open-shell molecules containing nuclei with spin $I \geq 1/2$ are also sensitive to nuclear-spin-dependent parity violating (NSD-PV) effects. In systems containing heavy nuclei, the NSD-PV effects are dominated by the P-odd anapole moment contribution, which arises from weak interactions within the nucleus, and couples to the electrons \cite{flambaum1980p,sushkov1978nature} (the notion of anapole moment of anelementary particle was introduced by Ya. B. Zel'dovich \cite{zel1958electromagnetic}). Therefore, measurements of the anapole moments can be used to test low energy quantum chromodynamics and P violation in nuclei \cite{GinFla04,SafBudDeM18}. So far, only one unambiguous measurement of the anapole moment was achieved in an experiment on the $^{133}$Cs atom \cite{wood1997measurement}, and a number of other atomic and molecular experiments are ongoing  \cite{AntFabBou17,AntFabSta19,aubin2013atomic,ChoEll16,DeMCahMur08,AltAmmCah18,Kar24}. 

The LaO, LaS, and LuO molecules are proposed as attractive candidates for measurements of the P- and the P, T-violating effects \cite{FlaKhr85}. First of all, these molecules contain heavy nuclei ($Z_{\mathrm{La}}=57$, $Z_{\mathrm{Lu}}=71$) and the P, T- and P- violating effects scale with powers of the atomic number $Z$. Additionally, these polar molecules with a $^2\Sigma$ ground state are well studied, stable in the gas phase, and possess a relatively simple electronic and rotational structure. Most importantly, they benefit from a partial cancellation of the hyperfine and rotational splittings, leading to very small intervals (less than 1 cm$^{-1}$ \cite{FlaKhr85}) between levels of opposite parity. These intervals may be reduced further by applying an external magnetic field, leading to full polarization of these molecules and significant enhancement of their sensitivity to the P and P, T violation. 

$^{139}$La has a nuclear spin $I=7/2$, and $^{175}$Lu and $^{176}$Lu have $I=7/2$ and $I=7$, respectively, making these nuclei suitable for investigation of both NMQM and anapole moment effects. Furthermore, the two isotopes of Lu are quadrupole deformed, and enhancement of the NMQM effects can be expected in these systems \cite{Fla94}. The NMQM of the Lu nucleus was calculated using the deformed oscillator Nilsson model for proton and neutron orbitals \cite{MaiSkrFla20,LacFla18}. La is a spherical nucleus and we calculate its NMQM here following \cite{SusFlaKhr84,FlaDeMKoz14}. The nuclear anapole moments of $^{139}$La and $^{175}$Lu isotopes were estimated in \cite{FlaKhr85}. 

The interpretation of the various experiments that search for P, T- and P-violating effects in molecular experiments and the extraction of the properties of interest from the measured energy shifts or transition amplitudes requires knowledge of electronic coupling parameters (also often referred to as enhancement factors). These coupling parameters are designated $W_{\mathrm{d}}$, $W_{\mathrm{s}}$, $W_{\mathrm{M}}$, and $W_{\mathrm{A}}$ for the eEDM, S-PS, NMQM, and NSD-PV anapole interactions, respectively. They can not be measured directly and have to be provided by theory. The accuracy and reliability of the calculated coupling parameters are important for the interpretation of the experiments, as is the knowledge of the uncertainty of the theoretical predictions. Thus, state-of-the-art computational methods that treat both relativistic effects and the electron correlation on a high level are employed.  

In this work, we calculate the $W_{\mathrm{d}}$, $W_{\mathrm{s}}$, $W_{\mathrm{M}}$, and $W_{\mathrm{A}}$ coefficients of the LaO, LaS, and LuO molecules using the relativistic coupled-cluster approach. Furthermore, we perform a computational study that allows us to set reliable error bars on our predictions, using the scheme developed in our earlier works \cite{ChaBorEli22,DenHaoEli20,HaaDoeBoe21}. Using the predicted values of the magnetic quadrupole and the anapole moments for the Lu and La nuclei, we estimate the expected measurable energy shift in terms of CP-violating parameters and the transition amplitude associated with the NSD-PV anapole moment. We hope these results can motivate experimental studies on these systems. Our calculations will be important for the planning and interpretation of future measurements. 
 
Some of the coupling constants of the LaO and LuO were calculated in the past, on varying levels of theory \cite{DeMCahMur08,BorIliDzu13,GauMarIsa19,MaiSkrFla20,ZhaZheChe21}. This work presents a complete study of the four parameters of the LaO, LaS, and LuO, on equal footing and includes reliable error estimates.\\

\section{Symmetry-violating molecular effective Hamiltonian}\label{sec:SymmetryViolations}
The P, T-violating eEDM operator can be expressed as \cite{LinLynSan89}
\begin{equation}
    H_{d_e}=2icd_e \sum^{N_{\mathrm{ele}}}_i  \gamma_i^0 \gamma_i^5\vv{p}_i^2,
\label{eq:H_de}    
\end{equation}
where $d_e$ represents the eEDM, $\gamma^5=i\gamma^0\gamma^1\gamma^2\gamma^3$, with $\gamma^0, \gamma^1, \gamma^2$, and $\gamma^3$, representing the Dirac gamma matrices, $c$ the speed of light, $\vv{p}_i$ is the momentum of electron $i$, and $N_{\mathrm{ele}}$ is the number of electrons.

The P, T-violating S-PS interaction due to the nucleus with atomic number $Z$ is \cite{GinFla04,ComDeM10}
\begin{equation}
   H_{k_s}=i \frac{G_F}{\sqrt{2}}Z k_{\mathrm{s}} \sum_i^{N_{\mathrm{ele}}} \gamma_i^0 \gamma_i^5\rho(\vv{r}_{i}),
   \label{eq:H_ks}
\end{equation}
where $k_{\mathrm{s}}$ represents the S-PS interaction strength, $G_F$ is the Fermi constant, $\rho(\vv{r}_{i})$ is the nuclear charge distribution, and $\vv{r}_i$ is the position of electron $i$ with respect to the nucleus. We assume here that the interaction is the same for both nucleons and we use $k_{\mathrm{s}}=(A/Z)C_s$ \cite{ComDeM10}, in contrast to $ZC_{s,p}+NC_{s,n}$, where the interactions with protons $C_{s,p}$, and neutrons $C_{s,n}$, are distinguished, and $A=Z+N$ \cite{GinFla04}. Here and in the following we consider the P, T
and P-- violating interactions only from the La and Lu heavy nuclei.

The P, T-violating NMQM interaction is given by \cite{SkrTitFla17}
\begin{equation}
    H_{M}=-\frac{M}{2I(2I-1)}T_{jk} \sum_i^{N_{\mathrm{ele}}}\frac{3}{2}\frac{[\vv{\alpha}_i \times \vv{r}_i]_j}{r_i^5}[r_i]_k,
    \label{eq:H_M}
\end{equation}
where $M$ is the NMQM, $T_{jk}$ are the components of the second-rank tensor $\vv{T}$, i.e. $T_{jk} = I_jI_k + I_kI_j - \frac{2}{3}I(I+1)\delta_{jk}$ with $I$ the spin of the nucleus. $\vv{\alpha}$ are the Dirac matrices.
The P, T--odd interaction of the electrons with the nuclear EDM produced by the Schiff moment in paramagnetic molecules is usually much smaller than the one produced by the NMQM \cite{KozLab95}. We included only the dominant NMQM interaction in the present work.

Finally, the NSD-PV interaction can be expressed as \cite{FlaKhr85}
\begin{equation}
    H_{\text{A}}=k_{\mathrm{NSD}} \frac{G_F}{\sqrt{2}}\sum_i^{N_{\mathrm{ele}}}\frac{\vv{\alpha}_i\cdot \vv{I}}{I}\rho(\vv{r}_i),
    \label{eq:H_a}
\end{equation}
where $k_{\mathrm{NSD}}=k_A+k_{2}+k_{Q}$ is the dimensionless strength constant. There are three sources for this interaction, with the nuclear anapole moment ($k_A$) being the dominant contribution for sufficiently large nuclear charge \cite{flambaum1980p,FlaKhrSus84}. The other two sources arise from the electroweak neutral coupling between electron vector and nucleon axial-vector currents ($\vv{V}_e \vv{A}_N$), $k_{2}$
\cite{NovSusFla77,nayak2009relativistic}, and from the nuclear-spin-independent weak interaction combined with the hyperfine interaction, $k_{Q}$ \cite{FlaKhr85Q}.

The effective Hamiltonian $H^{\text{eff}}$ for the P, T- and the P-violating interactions for the LaS, LaO, LuO diatomic molecules is given by \cite{KozLab95}
\begin{equation}\label{eq:H_eff}
    \begin{split}
        H^{\text{eff}}=&(d_e W_{\mathrm{d}} + k_{\mathrm{s}} W_{\mathrm{s}})\vv{J}\cdot \vv{{\hat{n}}}-\frac{M W_{\mathrm{M}}}{2 I(2 I-1)} \vv{J} \cdot{\vv{T}} \cdot\hat{\vv{n}}\\
        &+k_{\mathrm{NSD}} W_{\mathrm{A}} (\vv{\hat{n}}\times\vv{J})\cdot\frac{\vv{I}}{I},
    \end{split}
\end{equation}
where $\vv{J,I,\hat{n}}$ are the total electronic angular momentum, the nuclear spin and the unit vector along the internuclear axis, respectively.  

$W_{\mathrm{d}}$, $W_{\mathrm{s}}$, $W_{\mathrm{M}}$ represent the expectation value of the fundamental P, T-violating interactions \autoref{eq:H_de}, \autoref{eq:H_ks}, \autoref{eq:H_M}, respectively,
on the electronic molecular wave function of the $^2\Sigma_{\Omega=1/2}$ state, i.e.
\begin{equation}
        W_d=ic\frac{2}{\Omega}\bra{\Psi_{^2\Sigma_{1/2}}}
         \sum^{N_{\mathrm{ele}}}_i  \gamma_i^0 \gamma_i^5\vv{p}_i^2
        \ket{\Psi_{^2\Sigma_{1/2}}},
\end{equation}
\begin{equation}
        W_s=iZ\frac{G_F}{\sqrt{2}\Omega}\bra{\Psi_{^2\Sigma_{1/2}}}
        \sum_i^{N_{\mathrm{ele}}} \gamma_i^0 \gamma_i^5\rho(\vv{r}_{i})
        \ket{\Psi_{^2\Sigma_{1/2}}},
\end{equation}
\begin{equation}
        W_M=\frac{3}{2\Omega}\bra{\Psi_{^2\Sigma_{1/2}}}
        \sum_i^{N_{\mathrm{ele}}}\left(\frac{\vv{\alpha}_i \times \vv{r}_i}{r_i^5}\right)_\zeta r_\zeta
        \ket{\Psi_{^2\Sigma_{1/2}}},
\end{equation}

where $\Omega$ is the projection of the total electronic angular momentum $\vv{J}$ on the molecular axis, and $\zeta$ means projection on the molecular axis.
$W_{\mathrm{A}}$ represents the off-diagonal matrix element of \autoref{eq:H_a} between the electronic molecular states $^2\Sigma_{\Omega=1/2}$ and $^2\Sigma_{\Omega=-1/2}$,
\begin{equation}
        W_{\mathrm{A}}=\frac{G_F}{\sqrt{2}}\bra{\Psi_{^2\Sigma_{1/2}}}
        \sum_i^{N_\mathrm{ele}}\rho(\vv{r}_i)\alpha_+
        \ket{\Psi_{^2\Sigma_{-1/2}}},
\end{equation}

with $\alpha_+=\alpha_x + i\alpha_y$.

\section{Molecular enhancement factors}\label{sec:W-results}

We use the relativistic 4-component Dirac–Coulomb Hamiltonian combined with the single reference coupled-cluster approximation and the finite field approach. 

The unperturbed molecular Dirac--Coulomb Hamiltonian is given by
\begin{equation}
    H^{(0)}=\sum_i[\beta_imc^2 + c \vv{\alpha}_i \cdot \vv{p}_i-V_{\text{nuc}}(\vv{r}_i)],
\end{equation}
where $\vv{\alpha}_i$ and $\beta_i$ are the $4\times4$ Dirac matrices, $\vv{p}_i$ is the momentum of the electron $i$, and $V_{\text{nuc}}$ is the Coulomb potential energy at the position of the electron with respect to the considered nucleus $\vv{r}$. 

All the calculations were carried out using a modified version of the Dirac 2019 program \cite{DIRAC19,saue2020dirac}, and the uncontracted relativistic Dyall's basis sets \cite{gomes2010relativistic,dyall2010relativistic,dyall2016relativistic}. The experimental bond lengths used in the calculations of the molecular enhancement factors in the LaO and LuO molecules were 2.825~\AA\ \cite{konings2014thermodynamic} and 1.7902877~\AA\ \cite{cooke2011rotational}, respectively. 
The LaS bond length of 2.392~\AA\ was calculated in this work at the CCSD(T) level of theory with 30 frozen electrons and virtual space up to 30 a.u., using the dyall.v4z basis set.

\subsection{Computational approach}

The computational approach used in the calculation of the coupling molecular parameters is based on a systematic study of the effect of the electron correlation and the quality of the basis set on $W_{\mathrm{d}}$, $W_{\mathrm{s}}$, $W_{\mathrm{M}}$ and $W_{\mathrm{A}}$ in LaS and on $W_{\mathrm{d}}$ in LaO and LuO, as it is discussed in the following subsections. The final results are presented in \autoref{sec:Enhancements}.

\subsubsection{Electron correlation}
Calculations of the molecular enhancement factors at the coupled-cluster level of theory were shown to be highly sensitive to the number of electrons included in the correlation description \cite{ChaBorEli22,HaaDoeBoe21}. In this work, we correlated all electrons and use a virtual cut-off of 2000 a.u. in order to describe properly the low-lying occupied orbitals. We evaluated the missing electron correlation contributions from two sources: i) the incomplete virtual active space and ii) the truncation of the coupled-cluster expansion. The corresponding results are presented in \autoref{tab:correlation}. \\

i) The calculated molecular enhancement factors do not change considerably by including virtual orbitals over 2000 a.u., with $\sim0.3\%$ difference between including orbitals up to 3000 a.u. and up to 2000 a.u. and using the dyall.v3z basis set. Therefore, we included orbitals up to 2000 a.u. in all our calculations and considered the effect of neglecting higher-energy virtual orbitals as a source of error in our uncertainty estimation.\\

ii) The final results for the molecular enhancement factors are obtained on the coupled-cluster level including single, double and perturbative triple excitations, CCSD(T). The comparison of these results with the ones obtained with single and double excitations, CCSD, suggests that there is a non-negligible effect of higher-order excitations on the calculated enhancement factors, up to $4.2\%$ in LaS, LaO, and $5.8\%$ in LuO when using the cv3z basis set. 
We included the difference between the CCSD(T) and the CCSD results as an estimation of the missing higher-excitations effects, see \autoref{sec:uncertainty}. 

\begin{table}
\caption{Electron correlation effects on the calculated $W$ coupling parameters. Where not specified, the CCSD(T) level of theory was employed and virtual orbitals up to 2000 a.u. were correlated.}
\centering
\begin{tabular}{lrrrrrr}
\hline
 & \multicolumn{4}{c}{LaS} & LaO & LuO \\
 \cmidrule(lr){2-5} \cmidrule(lr){6-6} \cmidrule(lr){7-7} 
 Method & $W_\mathrm{d}$ & $W_\mathrm{s}$ & $W_\mathrm{M}$ & $W_\mathrm{A}$ & \multicolumn{2}{c}{$W_\mathrm{d}$} \\
& {[}$10^{24} \frac{h\text{ Hz}}{e\text{ cm}}${]} & {[}$h$ kHz{]} & {[}$10^{32} \frac{h\text{ Hz}}{e \text{ cm}^2}${]} & {[}$h$ Hz{]} & \multicolumn{2}{c} {[$10^{24} \frac{h\text{ Hz}}{e\text{ cm}}$]} \\
\hline
2000 a.u.  & 3.46 & 9.12 & 4.25 & 161.15 & 3.62 & 15.61 \\
3000 a.u.  & 3.47 & 9.14 & 4.26 & 161.59 & 3.63 & 15.61 \\
\hline
CCSD(T)& 3.65 & 9.63 & 4.46 & 170.22 & 3.81 & 15.67 \\
CCSD   & 3.79 &10.02 &	4.64 & 177.03 &	3.97 & 16.57 \\
\hline
\end{tabular}
\label{tab:correlation}
\end{table}

\subsubsection{Basis sets}
\begin{table}
\caption{Effect of diffuse (sv3z), core-correlating (cv3z, ae3z), and larger cardinality (v4z) dyall basis functions on the calculated $W$ coupling parameters.}
\centering
\begin{tabular}{lrrrrrr}
\hline
 & \multicolumn{4}{c}{LaS} & LaO & LuO \\
 \cmidrule(lr){2-5} \cmidrule(lr){6-6} \cmidrule(lr){7-7} 
 Basis & $W_\mathrm{d}$ & $W_\mathrm{s}$ & $W_\mathrm{M}$ & $W_\mathrm{A}$ & \multicolumn{2}{c}{$W_\mathrm{d}$} \\
& {[}$10^{24} \frac{h\text{ Hz}}{e\text{ cm}}${]} & {[}$h$ kHz{]} & {[}$10^{32} \frac{h\text{ Hz}}{e \text{ cm}^2}${]} & {[}$h$ Hz{]} & \multicolumn{2}{c} {[$10^{24} \frac{h\text{ Hz}}{e\text{ cm}}$]} \\
\hline
v2z        & 3.61 & 9.04 & 4.49 & 158.88 & 3.30 & 15.15 \\
v3z        & 3.46 & 9.12 & 4.25 & 161.15 & 3.62 & 15.61 \\
v4z        & 3.37 & 8.97 & 4.14 & 159.10 & 3.67 & --    \\
cv3z       & 3.65 & 9.63 & 4.46 & 170.22 & 3.81 & 15.67 \\
ae3z       & 3.72 & 9.83 & 4.54 & 173.83 & 3.88 & 15.67 \\
s-aug-v3z  & 3.45 & 9.09 & 4.23 & 160.63 & 3.64 & 15.61 \\
\hline
\end{tabular}
\label{tab:basis}
\end{table}

\begin{figure}
    \centering
    \includegraphics[scale=0.48]{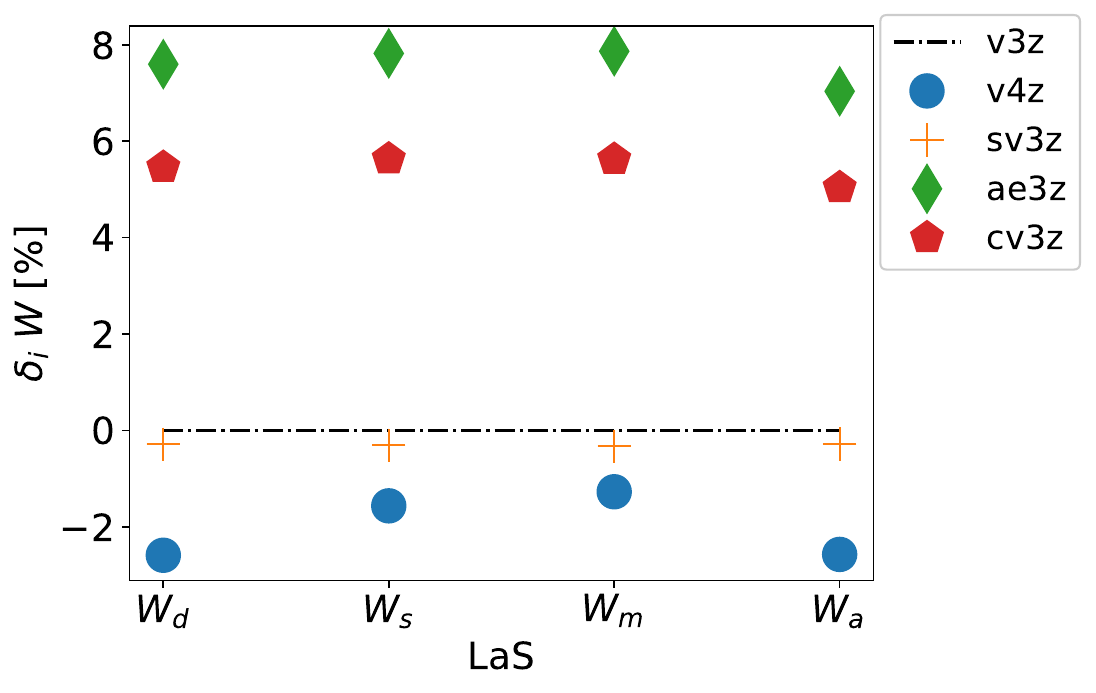}
    \includegraphics[scale=0.48]{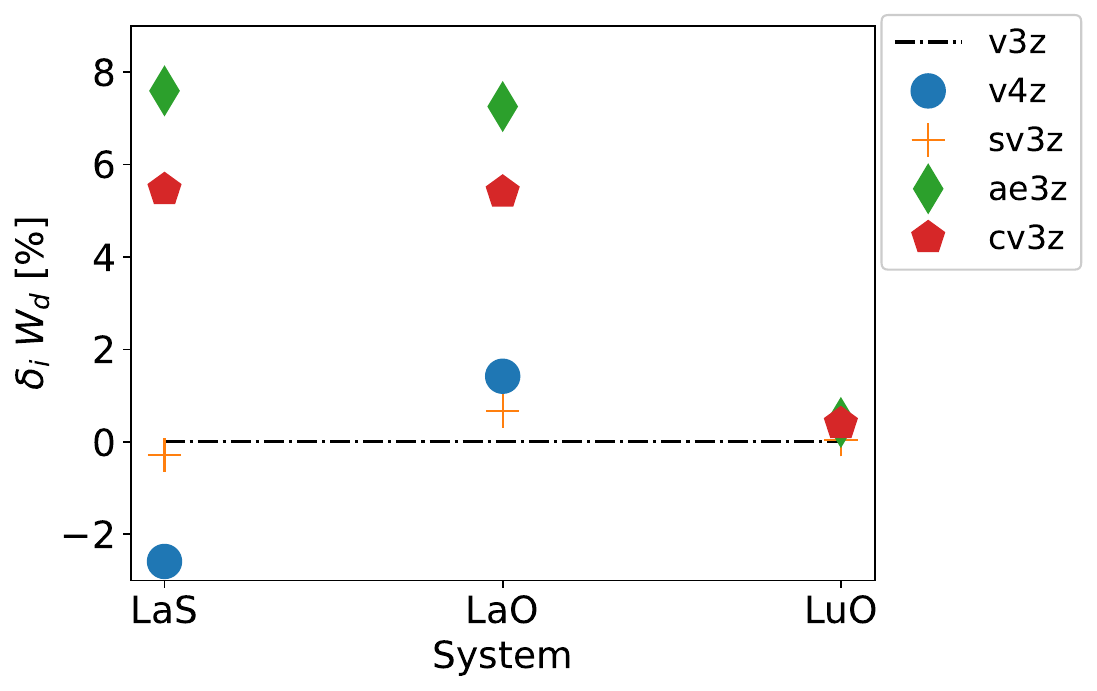}
    \caption{Effect of diffuse (sv3z), core-correlating (cv3z, ae3z), and larger cardinality (v4z) dyall basis functions on the calculated $W$ coupling parameters with respect to the values obtained with the dyall.v3z basis set (dashed line). $W_{\mathrm{d}}$, $W_{\mathrm{s}}$, $W_{\mathrm{M}}$, and $W_{\mathrm{A}}$ of LaS (top) and $W_{\mathrm{d}}$ of LaS, LaO and LuO (bottom).}
    \label{fig:basis-comparison}
\end{figure}

\begin{table}
\caption{Results obtained using the different CBS extrapolation schemes and their spread at 95\% confidence interval ($1.96\sigma$).}
\begin{tabular}{lrrrrrr}
 & \multicolumn{4}{c}{LaS} & LaO & LuO \\
 \cmidrule(lr){2-5} \cmidrule(lr){6-6} \cmidrule(lr){7-7} 
 & $W_\mathrm{d}$ & $W_\mathrm{s}$ & $W_\mathrm{M}$ & $W_\mathrm{A}$ & \multicolumn{2}{c}{$W_\mathrm{d}$} \\
& {[}$10^{24} \frac{h\text{ Hz}}{e\text{ cm}}${]} & {[}$h$ kHz{]} & {[}$10^{32} \frac{h\text{ Hz}}{e \text{ cm}^2}${]} & {[}$h$ Hz{]} & \multicolumn{2}{c} {[$10^{24} \frac{h\text{ Hz}}{e\text{ cm}}$]} \\
\hline
CBS(L) & 3.26 & 8.69 & 4.01 & 154.33 & 3.68 & 16.37 \\
CBS(H) & 3.28 & 8.76 & 4.03 & 155.33 & 3.68 & 16.36 \\
CBS(M) & 3.30 & 8.81 & 4.06 & 156.23 & 3.67 & 16.35 \\
\hline
95\% c.i. & 0.04 & 0.11 & 0.05 & 1.95 & 0.004 & 0.02 \\
\hline
\end{tabular}\label{tab:CBSE}
\end{table}

In this work, we used the Dyall basis sets \cite{gomes2010relativistic,dyall2010relativistic,dyall2016relativistic}, which are uncontracted Gaussian functions optimized for relativistic calculations. Using these basis sets, we studied the variation of the molecular enhancement factors with the addition of diffuse and core-correlating functions, and with the cardinality. In these calculations, we correlated all the electrons and included virtual orbitals up to 2000 a.u. on the CCSD(T) level of theory, see \autoref{tab:basis}.
\autoref{fig:basis-comparison} shows the effect of the basis set on the calculated enhancement factors. The top panel presents the effect on the 4 constants in the same molecule, LaS, while the bottom panel focuses on the $W_{\mathrm{d}}$ parameter in the three molecules under study.  In both cases the results are not sensitive to the addition of diffuse functions, while the core-correlating functions have a significant effect. The P, T-- and P--, violating effects involve nuclear mechanisms and therefore it can be expected that high-quality calculations of molecular enhancement factors require an accurate description of the electronic environment surrounding the nuclei, for which the presence of core-correlating functions is necessary. Note that there is a larger difference between the dyall.v3z and dyall.cv3z results than between dyall.v3z and dyall.v4z; this shows that the effect of core-correlating functions is more significant than that of basis set cardinality. Due to the high computational costs, we did not include the dyall.v4z basis set result for LuO in \autoref{fig:basis-comparison}, but refer the reader to \autoref{tab:cardinality-LuO}, where the same tendency is observed when correlating a smaller number of electrons. Optimising accuracy and realistic computational requirements, we used the dyall.cv3z basis set for our final results. 

Furthermore, we extrapolated our results to the complete basis set limit (CBSL), using the v$n$z basis set ($n=2,3,4$). We used the three-point Dunning--Feller $e^{-\alpha n}$ scheme \cite{dunning1989gaussian, feller1992application} for extrapolating the DHF energies and the two-point Helgaker et al. $n^{-3}$ scheme ($n=3,4$) \cite{helgaker1997basis} for extrapolating the correlation energies. We also tested the Martin $(n+\frac{1}{2})^{-4}$ scheme \cite{martin1996ab} and the scheme of Lesiuk and Jeziorski \cite{lesiuk2019complete} for extrapolating the correlation energies. We found that the three schemes (Lesiuk L, Helgaker H, and Martin M) give very similar CBS limits, see \autoref{tab:CBSE}. We used the difference CBS(H) - v3z to estimate the uncertainty in the final results due to the basis set incompleteness.

\subsubsection{Uncertainty estimation}\label{sec:uncertainty} 

\begin{table*}
\caption{Summary of the contributions of the investigated sources of uncertainty in the calculation of the molecular enhancement factors. The relative percentage to the final $W$ value is presented in parenthesis.}
\begin{tabular}{lrrrrrrrrrrrrl}
\hline
 & \multicolumn{8}{c}{LaS} & \multicolumn{2}{c}{LaO} & \multicolumn{2}{c}{LuO} &  \\
 \cmidrule(lr){2-9} \cmidrule(lr){10-11} \cmidrule(lr){12-13}
{Source} & \multicolumn{2}{c}{$\delta_i W_\mathrm{d}$} & \multicolumn{2}{c}{$\delta_i W_\mathrm{s}$} & \multicolumn{2}{c}{$\delta_i W_\mathrm{M}$} & \multicolumn{2}{c}{$\delta_i W_\mathrm{A}$} & \multicolumn{2}{c}{$\delta_i W_\mathrm{d}$} & \multicolumn{2}{c}{$\delta_i W_\mathrm{d}$} & Scheme \\

 & \multicolumn{2}{c}{{[}$10^{24} \frac{h\text{ Hz}}{e\text{ cm}}${]}} & \multicolumn{2}{c}{{[}$h$ kHz{]}} & \multicolumn{2}{c}{{[}$10^{32} \frac{h\text{ Hz}}{e \text{ cm}^2}${]}}  & \multicolumn{2}{c}{{[}$h$ Hz{]}} & \multicolumn{2}{c}{{[}$10^{24} \frac{h\text{ Hz}}{e\text{ cm}}]$} & \multicolumn{2}{c}{{[}$10^{24} \frac{h\text{ Hz}}{e\text{ cm}}]$} &  \\
 \hline
CBSE            & 0.18&(4.9\%)& 0.36&(3.7\%)& 0.21&(4.7\%)& 5.82&(3.4\%)& 0.06&(1.7\%)& 0.10&(0.6\%) & CBS(H) -- v3z\\
Valence         & 0.01&(0.3\%)& 0.03&(0.3\%)& 0.01&(0.3\%)& 0.53&(0.3\%)& 0.02&(0.6\%)& 0.01&(0.0\%) & s-v3z -- v3z\\
Core            & 0.07&(2.0\%)& 0.20&(2.1\%)& 0.08&(1.9\%)& 3.61&(2.1\%)& 0.07&(1.8\%)& 0.00&(0.0\%) & ae3z -- cv3z\\
\hline
Virt. cutoff    & 0.01&(0.3\%)& 0.03&(0.3\%)& 0.01&(0.3\%)& 0.43&(0.3\%)& 0.01&(0.3\%)& 0.01&(0.1\%) & 3000 -- 2000 a.u.\\
Higher exc.     & 0.15&(4.0\%)& 0.39&(4.0\%)& 0.18&(4.0\%)& 6.81&(4.0\%)& 0.16&(4.2\%)& 0.91&(5.8\%) & CCSD(T) -- CCSD\\
\hline
Geometry        & 0.09&(2.4\%)& 0.23&(2.4\%)& 0.11&(2.4\%)& 4.06&(2.4\%)&--&&--& & variation in 0.05 \AA\\
\hline
Total           & 0.26&(7.1\%)& 0.61&(6.3\%)& 0.31&(6.9\%)&10.50&(6.2\%)& 0.19&(4.9\%)& 0.91&(5.8\%) & $\sqrt{\sum_i \delta_i^2}$\\
Final value     & 3.65&& 9.63&& 4.46&& 170.22&& 3.81&& 15.67& &cv3z\\
\hline \hline
\end{tabular}\label{tab:uncertainty}
\end{table*}

\begin{figure}
    \centering
    \includegraphics[scale=0.6]{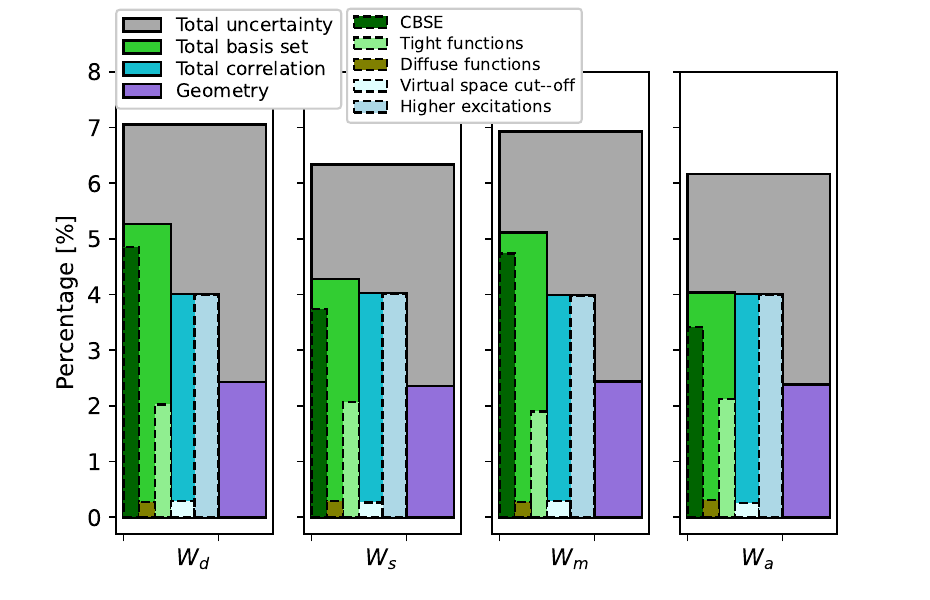}
    \includegraphics[scale=0.6]{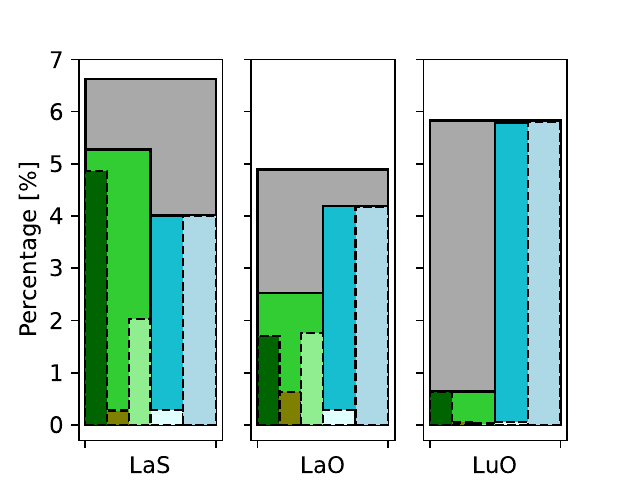}
    \caption{Scheme of the total and individual contributions to the uncertainty of all the calculated coupling parameters in LaS (upper) and of $W_{\mathrm{d}}$ in LaO, LaS and LuO molecules (bottom).}
    \label{fig:uncertainty}
\end{figure}

Based on our investigation of the electron correlation and basis set effects, we assessed the uncertainty in our calculated enhancement parameters. All the individual contributions and the total uncertainty, obtained assuming the effects are independent, are presented in \autoref{tab:uncertainty} and \autoref{fig:uncertainty}. 
The magnitude of each source of uncertainty corresponds to the difference in the enhancement factor obtained with the best and second-best approximation for a given computational parameter. In addition to the effect of the electron correlation and basis set, we include the uncertainty from the optimized geometry (LaS only) by calculating the effect of a variation of 0.05 \AA~ in the bond length on the enhancement factor \autoref{tab:geometry}. We used 0.05 \AA~ since this corresponds to the reported experimental range of the bond length in LaS \cite{marcano1970analysis}. The values used in the uncertainty estimation are presented in the previous section, \autoref{tab:correlation},\autoref{tab:basis},\autoref{tab:CBSE}. 

We focus our analysis on the LaS molecule since it is lighter than LuO, and therefore, the corresponding calculations are less expensive computationally. Columns 2--5 in \autoref{tab:uncertainty} and \autoref{fig:uncertainty} show that the relative contribution of uncertainty from each source, $\delta_i$, is approximately of the same size for all the calculated molecular enhancement factors, i.e. $\delta_i W_{\mathrm{d}} \sim \delta_i W_{\mathrm{s}} \sim \delta_i W_{\mathrm{M}} \sim \delta_i W_{\mathrm{A}}$ for each $i$. The difference in the overall uncertainty between $\delta W_{\mathrm{s}} \sim \delta W_{\mathrm{A}}$ ($\sim6$\%) and $\delta W_{\mathrm{d}} \sim \delta W_{\mathrm{M}}$ ($\sim$7\%) comes from a smaller dependency of the first two parameters on the cardinality of the basis set.

Furthermore, considering that the effect of the computational parameters on a given property may be sensitive to the molecular identity, we included in our analysis the study of $\delta_i W_{\mathrm{d}}$ in LaS, LaO and LuO (see columns 2, 6, 7 in \autoref{tab:uncertainty} and \autoref{fig:uncertainty}). The overall uncertainties $\delta W_{\mathrm{d}}$ are up to $\sim7\%$; with two clear differences i) The contribition of the triple excitations to the uncertainty in LaO and LaS is similar and smaller than in LuO, and ii) the uncertainty due to the basis set incompleteness behaves as LaS $>$ LaO $>$ LuO. Overall, when all electrons are correlated and the core valence basis sets are used, the main contributions to the uncertainty across the properties and the molecules are the basis set cardinality and the neglect of the higher excitations, which provide a similar-sized contribution.
Based on our conclusions concerning the similarity of the relative error size across the different enhancement parameters for the same molecule, we estimated $\delta W_{\mathrm{s,M,A}} = \delta W_{\mathrm{d}}$ for LaO and LuO. 

\subsection{Final enhancement factors}\label{sec:Enhancements}

\autoref{tab:enhancement} presents our final results and their corresponding uncertainties for the P, T-- and P--violating enhancement factors in LaO, LaS, and LuO. 
These enhancement factors are between 3 and 5 times larger in LuO than in LaO and LaS, due to the higher atomic number of Lu, while  LaO and LaS have very close values due to the negligible influence of the ligand.
We report small uncertainties of below 7\%.
We found that where available, previous calculations at the semiempirical \cite{DeMCahMur08}, 4c-DFT \cite{BorIliDzu13}, ZORA-HF and ZORA-DFT \cite{GauMarIsa19}, and X2C-CCSD(T) \cite{ZhaZheChe21,MaiSkrFla20} level are in agreement with our results, see \autoref{tab:enhancement}. 

\begin{table}
    \caption{Molecular enhancement factors and associated uncertainties on the LaS, LaO and LuO molecules calculated in this work. Previously reported values are shown for comparison.}
    \centering  
   \begin{tabular}{lrrrrr}
    \hline
     & $W_\mathrm{d}$ & $W_\mathrm{s}$ & $W_\mathrm{M}$ & $W_\mathrm{A}$ & Method\\
    & {[}$10^{24} \frac{h\text{ Hz}}{e\text{ cm}}${]} & {[}$h$ kHz{]} & {[}$10^{32} \frac{h\text{ Hz}}{e \text{ cm}^2}${]} & {[}$h$ Hz{]} \\
    \hline
    LaS & 3.65(26) &  9.63(61) &  4.46(31) &  170(11) & CCSD(T)\\
    LaO & 3.81(19) & 10.25(50) &  4.55(22) &  182( 9) & CCSD(T)\\
        & 3.71/4.76 & 10.1/13.0 & -- & -- & HF/DFT\cite{GauMarIsa19}\\
        &--&--&--&180.2& DFT\cite{BorIliDzu13}\\
        &--&--&--& 222  & Semiemp. \cite{DeMCahMur08}\\
    LuO &15.7(9)   & 57(3)     & 14.3(8)   &   814(47)& CCSD(T)\\
        & 15.2/17.9 & 55.9/65.7 & -- & -- & HF/DFT\cite{GauMarIsa19}\\
        & 15.7 &--&--&--&CCSD(T)\cite{ZhaZheChe21}\\
    \hline \hline
    \end{tabular}
    \label{tab:enhancement}
\end{table}

\section{Energy shifts and matrix elements}\label{sec:energy} 

\subsection{Parity and time-reversal violation}

The effective Hamiltonian describing the electron electric dipole moment in \autoref{eq:H_eff} leads to an energy shift $\Delta E$ for a state $\ket{\Psi}$ in an electric field $\mathcal{E}$,
\begin{equation}
   \Delta E = d_e W_{\mathrm{d}} \bra{\Psi}\vv{J} \cdot \vv{\hat{n}} \ket{\Psi}.
\end{equation}
Here $\bra{\Psi}\vv{J} \cdot \vv{\hat{n}} \ket{\Psi}=\Omega P(\mathcal{E})$, where the polarization factor $P(\mathcal{E})$ describes the mixing of opposite parity eigenstates in the electric field, and $\Omega$ is the projection of the total electronic angular momentum on the molecular axis \cite{boe23}. We therefore have
\begin{equation}
    \Delta E =d_e W_{\mathrm{d}} P(\mathcal{E}) \Omega.
\end{equation}

Similarly, for the scalar-pseudoscalar nucleon-electron interaction, the energy shift is
\begin{equation}
    \Delta E = k_{\mathrm{s}} W_{\mathrm{s}} P(\mathcal{E}) \Omega.
\end{equation}
The eEDM and the S-PS interaction constants, $d_e$ and $k_{\mathrm{s}}$ have a contribution from the Standard Model, i.e. from the complex phase in the CKM matrix. The estimated values are  \cite{YamYam20,EmaGaoPos22}
\begin{equation}
    d_e ^{\rm{CKM}}\sim 5.8 \times 10^{-40} e~\mathrm{cm} ,
\end{equation}
\begin{equation}
    k_{\mathrm{s}} ^{\rm{CKM}}\sim 6.9 \times 10^{-16}.
\end{equation}

Since the experimental limits on $d_e$ and $k_{\mathrm{s}}$ are $|d_e| <4.1 \times 10^{-30}$ and $|k_{\mathrm{s}}| < 1.4 \times 10^{-10}$ \cite{RouCalWri23}, there is a large window to find new physics by searching for these interactions. 
$k_{\mathrm{s}}$ is also sensitive to hadronic CP violation, 
parametrized by the QCD vacuum angle $\theta$, 
\begin{equation}
    k_{\mathrm{s}}(\theta) \approx 0.03\theta,
\end{equation}
which shows that EDMs of paramagnetic molecules also can restrict $\theta$ and may be used to search for axion dark matter (which gives oscillating   $\theta$) \cite{FlaPosRit20,FlaSamTan20l,FlaSamTan20,RosPalCai21}.\\

The effective Hamiltonian describing the NMQM in \autoref{eq:H_eff}, for the maximum nuclear spin projection $I=I_z$ along $\vv{\hat{n}}$, leads to an energy shift \cite{HoLimSau23} 
\begin{equation}\label{eq:deltaE-M}
    \Delta E = M W_{\mathrm{M}} \xi P(\mathcal{E}).
\end{equation}
$\xi$ depends on the nuclear and electronic spin projections and $P(\mathcal{E})$ is again the polarization factor.

The NMQM $M$ can be calculated for a spherical nucleus following \cite{SusFlaKhr84,FlaDeMKoz14}. Particularly, for $^{139}$La, 
\begin{equation}
    M^{\mathrm{La}}=3.5 \eta_p \times 10^{-34} e~\mathrm{cm}^2 - 1.0 d_p \times 10^{-13}~ \mathrm{cm},
\end{equation}
where $\eta_p$ and $d_p$ represent the proton-nucleus P,T-V strength constant and the proton EDM, respectively.

In a deformed nucleus, each nucleon from external nuclear shell  gives $M=4\Sigma \Lambda M_0^{\nu}$, where $\Sigma$ and $\Lambda$ are projections of the nucleon spin and orbital angular momentum on the nuclear axis, and $M_0^{\nu}$ are the single particle matrix elements for protons ($p$) and neutrons ($n$) \cite{LacFla18}. Summation over nucleons in $^{175}$Lu gives the collective NMQM \cite{MaiSkrFla20}
\begin{equation}
    M^{\mathrm{Lu}} = 15M^p_0 + 32M^n_0.
\end{equation}
 
Using the single particle matrix elements 
$M_0^p = -0.76 \eta_p \times 10^{-34} e~\mathrm{cm}^2 + 2.1 d_p \times 10^{-14}\mathrm{cm}$, and 
$M_0^n = 0.80 \eta_n \times 10^{-34} e~\mathrm{cm}^2 + 2.1 d_n \times 10^{-14}\mathrm{cm}$ 
\cite{FlaDeMKoz14,LacFla18}, we obtain
\begin{equation}
    \begin{split}
            M^{\mathrm{Lu}} = & - 1.14 \eta_p \times 10^{-33} \mathrm{cm}^2 e + 3.15 d_p \times 10^{-13} \mathrm{cm} \\
            &+2.56 \eta_n \times10^{-33} \mathrm{cm}^2 e  + 6.72  d_n \times 10^{-13} \mathrm{cm}.
    \end{split}
\end{equation}
The P-,T-violating nuclear potential is dominated by the neutral pion exchange \cite{MaiSkrFla20}. The constants $\eta_{\nu}$ can be expressed in terms of the $\pi NN$ strong coupling constant $g$, and P, T-violating $\pi NN$ coupling constants in different isospin channels $\bar{g}_i; i=0-2$ and
$\eta_n = -\eta_p \approx 5 \times 10^6 g (\bar{g}_1 + 0.4\bar{g}_2 -0.2\bar{g}_0)$ \cite{FlaDeMKoz14}. Furthermore, the $\pi NN$ constants can be expressed in terms of fundamental P, T-violating parameters, the QCD vacuum angle $\theta$, or EDMs $d_{u,d}$ and chromo-EDMs $\Tilde{d}_{u,d}$ of the up and down quarks \cite{YamSahYosh17},

\begin{equation}
    \begin{split}
        g\bar{g}_0(\theta)&= -0.2\bar{\theta}\\
        g\bar{g}_1(\theta)&= 0.046\bar{\theta}\\
        d_n&= -d_p = 1.2 \times10^{-16} \bar{\theta}~e~\mathrm{cm}\\
        g\bar{g}_0(\Tilde{d}_u, \Tilde{d}_d)&= 0.8 \times 10^{15}(\Tilde{d}_u + \Tilde{d}_d) \mathrm{cm}^{-1}\\
        g\bar{g}_1(\Tilde{d}_u, \Tilde{d}_d)&= 4 \times 10^{15}(\Tilde{d}_u - \Tilde{d}_d) \mathrm{cm}^{-1}\\
        d_p(d_u,d_d,\Tilde{d_u},\Tilde{d_d})&= 1.1 e (\Tilde{d}_u + 0.5\Tilde{d}_d) + 0.8 d_u - 0.2 d_d\\
        d_n(d_u,d_d,\Tilde{d_u},\Tilde{d_d})&= 1.1 e (\Tilde{d}_d + 0.5\Tilde{d}_u) + 0.8 d_d - 0.2 d_u.
    \end{split}
\end{equation}

\autoref{tab:nMQM} presents the energy shift \autoref{eq:deltaE-M} in terms of the discussed fundamental constants using $\xi P(\mathcal{E})=1$.

\begin{table}
\caption{Molecular enhancement factors and dependency of $M$ on CP violating parameters $x$.\label{tab:nMQM}}
\begin{tabular}{lrrrrr}
\hline
 &  & \multicolumn{4}{c}{$\partial W_{\mathrm{M}} M / \partial x$ ~(Hz)} \\
System & $W_{\mathrm{M}}$ & $d_p$ & $\theta$ & $\tilde{d}_d$ & $\tilde{d}_u$ \\
 & $10^{32}h\mathrm{Hz} /e\mathrm{cm}^2$ & $10^{20}e/ \mathrm{cm}$ & $10^4$ & $10^{21}/\mathrm{cm}$ & $10^{21}/\mathrm{cm}$ \\
 \hline
LaS & 4.24 & 5.02 & --6.53 & 3.06 & --2.90 \\
LaO & 4.63 & 5.48 & --7.13 & 3.35 & --3.16 \\
LuO & 14.37 & --200.08 & 245.53 & --109.28 & 103.11\\
\hline \hline
\end{tabular}
\end{table}

\subsection{Nuclear spin-dependent parity violation}

The effective Hamiltonian describing the NSD-PV interaction in \autoref{eq:H_eff} has non-zero matrix elements $me$ between opposite parity spin rotational states $\ket{\Psi^{+}(m_F)}$ and $\ket{\Psi^{-}(m_F')}$ if $m_F=m_F'$, where $m_F=m_N+m_I+m_S$. Here $m_F$, $m_N$, $m_I$, and $m_S$ are the projections of the total (F), rotational (N), nuclear spin (I), and electronic spin (S) angular momentum on the quantization axis, respectively  \cite{DeMCahMur08}. In particular, 
\begin{equation}\label{eq:ME-PV}
    me = k_{\mathrm{NSD}} W_{\mathrm{A}} C_A, 
\end{equation}
with $C_A=\bra{\Psi^-(m_F)}(\vv{n}\times \vv{J}) \cdot \vv{I}/I\ket{\Psi^+(m_F)}$.

The dimensionless constant $k_{\rm{NSD}}$ is composed of three contributions, $k_{\mathrm{NSD}}=k_a+k_2+k_Q$. $k_a$ is proportional to the anapole moment, $k_2$ to the $V_eA_n$ term in the electron-nucleus Z$^0$-exchange, and $k_Q$ to the weak interaction from the nuclear charge perturbed by the hyperfine interaction. $k_a$ and $k_2$ have been estimated from a nuclear shell model consisting of a single valence nucleon around a uniform core as \cite{FlaKhrSus84}
\begin{equation}
    k_a = 1.0 \times 10^{-3} g_{\nu} \mu_{\nu} A^{2/3} \frac{K}{I+1},
\end{equation}
where $g_{\nu}$ is the PV interaction strength between the core and the nucleon $\nu$ (we use $g_{p}=4.5$), $\mu_{\nu}$ is the nucleon magnetic moment, and $K=(I+1/2)(-1)^{I+1/2-\ell}$, with the nucleon orbital quantum number $\ell$, and
\begin{equation}
     k_2 = C_{2\nu} \frac{1/2-K}{I+1},
\end{equation}
with $C_{2\nu}\approx0.05$ \cite{GinFla04}. $k_Q$ has been estimated in terms of the mass number and the magnetic moment of the nucleus $\mu_N$ in nuclear magnetons as \cite{FlaKhr85Q},
\begin{equation}
    k_Q = 2.5 \times 10^{-4} A^{2/3} \mu_N.
\end{equation}

Using the previous expressions, we estimate $k_{\mathrm{NSD}}$ for LaO, LaS and LuO in \autoref{tab:anapole}. Note that the $^{175}$Lu nucleus is deformed. 
The matrix element $C_A$ in \autoref{eq:ME-PV} can be calculated using angular momentum algebra, as presented in \cite{Rah10}. For a given system, this matrix element depends on the mixing of the molecular states in applied electric and magnetic field. \autoref{tab:anapole} presents our results in terms of $C_A$. For the LaO molecule, the matrix element at the level crossing with maximum value of $m_F$ has been calculated in reference ($C_A=0.43$) \cite{DeMCahMur08}, which we include here.

\begin{table}
        \caption{Strength constants, molecular enhancement factor and matrix elements for the NSD-PV interaction.}
        \centering
        \begin{tabular}{ccrrrrrrrr}
        \hline
        System  & Nucleus & I & 100$k_a$ & 100$k_2$ & 100$k_Q$ & $W_{\mathrm{A}}$ [Hz] & $me$ [Hz]  \\
        \hline
        LaS & $^{139}$La$_{57}$ & 7/2 & 30.0 & --3.9 & 1.8 & 170 &  20\\
        LaO & $^{139}$La$_{57}$ & 7/2 &   30.0 & --3.9 & 1.8 & 182 & 22$C_A$\\
        LuO & $^{175}$Lu$_{71}$ & 7/2 & 34.9 & --3.9 & 1.7 &  814 & 266$C_A$ \\ 
        \hline
        \end{tabular}
        \label{tab:anapole}
\end{table}

\section{Conclusions}

We calculated the molecular enhancement factors for the P, T- violating electron electric dipole moment ($W_{\mathrm{d}}$), scalar-pseudoscalar nucleon-electron interaction ($W_{\mathrm{s}}$), and nuclear magnetic quadrupole moment ($W_{\mathrm{M}}$), and for the nuclear spin-dependent parity violating anapole moment ($W_{\mathrm{A}}$) in the LaS, LaO and LuO molecules. We estimated the uncertainties in the calculated values to be up to $7\%$. We used our calculated enhancement factors and nuclear models to estimate the energy shifts due to the NMQM in terms of fundamental constants, and the matrix element associated to the nuclear spin-dependent parity violating anapole moment in the studied molecules. 

Experiments on the systems considered here would add valuable information for probing different sources of P, T- violating effects since they are, beside the eEDM and S-PS interactions, also sensitive to the NMQM due to a nuclear spin $I>1$ \cite{GauBer23}. Additionally, the ratio of $W_s/W_{\mathrm{d}}$ in the LuO (3.65) differs considerably from that in the most sensitive experiments currently performed ($\sim1.7-2.8$). Therefore a competitive limit obtained in this system would place tighter constraints the P,T-violating parameter space \cite{GauMarIsa19,GauBer23}. 

\section{Acknowledgments}
We thank the Center for Information Technology at the University of Groningen for their support and for providing access to the Peregrine and Hábrók high-performance computing clusters. 
LFP acknowledges the support from the Dutch Research Council (NWO) project number VI.C.212.016 of the talent programme VICI, and the support from the Slovak Research and Development Agency (projects APVV-20-0098, APVV-20-0127). VVF acknowledges the support from the Australian Research Council Grants No. DP230101058 and DP200100150.
RFGR acknowledges the support from the U.S. Department of Energy, Office of Science, Office of Nuclear Physics under the grants DE-SC0021176 and DE-SC0021179.

\section{Supplementary information}\label{sec:SI}

We estimated the effect of the enhancement factors coming from a variation of 0.5 \AA~ in the bond length. \autoref{tab:geometry} presents values used in the uncertainty estimation.

\begin{table}[h!]
\caption{Effect of the variation of bond length of the molecular enhancement factors is LaS. }
\begin{tabular}{lllll}
\hline
 Bond length & $W_\mathrm{d}$ & $W_\mathrm{s}$ & $W_\mathrm{m}$ & $W_\mathrm{a}$ \\
 & {[}$10^{24} \frac{h\text{ Hz}}{e\text{ cm}}${]} & {[}$h$ kHz{]} & {[}$10^{32} \frac{h\text{ Hz}}{e \text{ cm}^2}${]} & {[}$h$ Hz{]}  \\
\hline
2.36 & 3.59 & 9.48 & 4.39 & 209.13 \\
2.42 & 3.68 & 9.71 & 4.50 & 214.19 \\
\hline \hline
\end{tabular}\label{tab:geometry}
\end{table}

We use the results presented in \autoref{tab:basis} lines 1--3 to calculate the CBS limit on the enhancement parameters. To reduce the computational cost, 10 electrons were frozen and the CCSD level of theory was employed in the calculations of $W_d$ for LuO with the dyall.v$n$z, $n=2-4$ basis set, see the results in \autoref{tab:cardinality-LuO}.

\begin{table}[h!]
\caption{Values used for the CBS extrapolation for $W_d$ in LuO.}
\centering
\begin{tabular}{ll}
\hline
Basis set & CCSD \\
\hline
v2z &15.85 \\
v3z &16.26 \\
v4z &16.31\\
\hline \hline
\end{tabular}\label{tab:cardinality-LuO}
\end{table}

%
\providecommand{\noopsort}[1]{}\providecommand{\singleletter}[1]{#1}%
\end{document}